\documentclass[aps,prl,twocolumn,showpacs,floatfix]{revtex4}

\usepackage{graphics}
\usepackage{epsfig}
\usepackage{amsfonts}

\begin{document}

\title{Relevance of ferromagnetic correlations for the Electron Spin Resonance in Kondo lattice systems}
\author{C. Krellner}
\author{T. F\"orster}
\author{H. Jeevan}
\author{C. Geibel}
\author{J. Sichelschmidt}

\affiliation{Max Planck Institute for Chemical Physics of Solids, D-01187 Dresden, Germany}

\date{\today}

\begin{abstract}
Electron Spin Resonance (ESR) measurements of the ferromagnetic Kondo lattice system CeRuPO show a well defined ESR signal which is related to the magnetic properties of the Ce$^{3+}$ moment. In contrast, no ESR signal could be observed in the antiferromagnetic homologue CeOsPO. Additionally, we detect an ESR signal in a further ferromagnetic Yb compound, YbRh, while it was absent in a number of Ce or Yb intermetallic compounds with dominant antiferromagnetic exchange, independently of the presence of a strong Kondo interaction or the proximity to a (quantum) critical point. Thus, the observation of an ESR signal in a Kondo lattice is neither specific to Yb nor to the proximity of a quantum critical point, but seems to be connected to the presence of ferromagnetic fluctuations. These conclusions not only provide a basic concept to understand the ESR in Kondo lattice systems even well below the Kondo temperature as observed in the heavy fermion metal $\mathrm{Yb Rh_2 Si_2}$ but point out ESR as a prime method to investigate directly the spin dynamics of the Kondo ion.
\end{abstract}


\maketitle
\begin{figure}[tbp]
  \centering
 \includegraphics[width=0.45\textwidth]{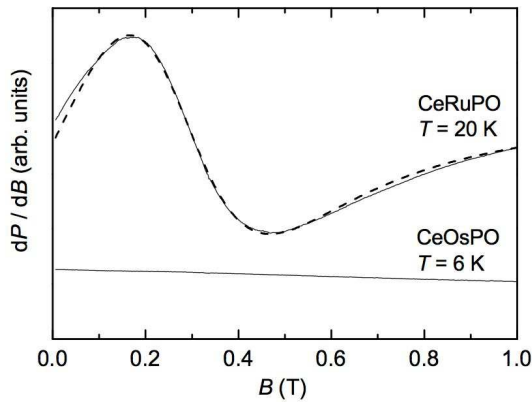}
  \caption{ESR spectra of polycrystalline samples of CeRuPO ($T_{\rm C}= 15$~K) and CeOsPO ($T_{\rm N}=4.4$~K). Dashed line represents a fit to the data with a metallic Lorentzian line shape at $g=2.6$. Spectra were taken with same instrument settings.}
 \label{fig1}
\end{figure}
\begin{figure}[tbp]
	  \centering
	 \includegraphics[width=0.45\textwidth]{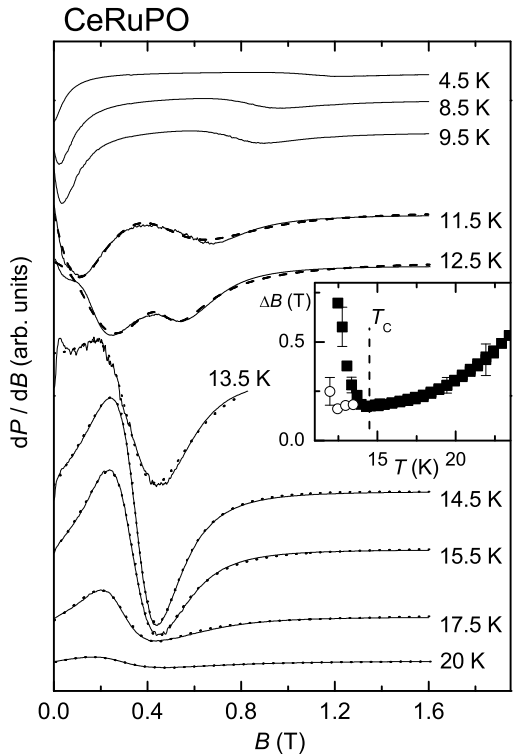}
	  \caption{ESR spectra of powder samples of CeRuPO below and above $T_{\rm C}= 15$~K. Dotted lines represent metallic Lorentzian line fits. Below $T_{\rm C}$ a low field and a high field component of the ESR line develops as shown with two Lorentzian lines (dashed line). Inset: $T$-dependence of the ESR linewidth $\Delta B$ (squares). The open circles refer to the high field component which appears below $T_{\rm C}$. }
	 \label{fig2}
	\end{figure}

%
%
Heavy fermion intermetallic compounds with concentrated 4$f$- and 5$f$- ions show no Electron Spin Resonance (ESR) signal. This common belief was experimentally well established and was justified by the strong electronic correlations which arise from the hybridization between 4$f$- or 5$f$-electrons and conduction electrons. An ESR investigation of the Kondo ions' spin dynamics seemed only to be possible indirectly by additionally doped ESR probe spins, usually Gd$^{3+}$ \cite{elschner97a,krug03a}, analogous to the principle of NMR in these systems. However, the dense Kondo lattice systems YbRh$_{2}$Si$_{2}$ and YbIr$_{2}$Si$_{2}$ pose a remarkable exception. They exhibit an ESR signal with clear properties of local Yb$^{3+}$ moments and, most surprisingly, the signal is observable below the Kondo temperature $T_{\rm K}$ where the local Yb$^{3+}$ moments become screened because of pronounced Kondo interactions.\cite{sichelschmidt03a,sichelschmidt07a} As both compounds are located very close to a quantum critical point \cite{gegenwart02a,hossain05a} the ESR signal was initially thought to be a direct verification of the localized moment scenario of quantum criticality \cite{sichelschmidt03a}. However, the consequences of this signal for understanding the physics close to a quantum critical point in particular or for the Kondo ion physics in general remained unclear. A plausible explanation would be the inherent difference between Yb- and Ce- based Kondo lattices, the former one presenting a much stronger local character allowing for a narrow observable ESR line. Another mechanism could be based on the presence of ferromagnetic fluctuations, since YbRh$_{2}$Si$_{2}$ and YbIr$_{2}$Si$_{2}$ seem to be unique cases of Kondo lattice systems close to a critical point with strong ferromagnetic fluctuations and a concomitant strongly enhanced spin susceptibility. Such a scenario is supported by previous ESR reports about how exchange enhanced spin susceptibility leads to a reduced ESR line broadening \cite{peter67a} compared to the line broadening in simple ferromagnetic systems like Fe or Ni \cite{cooper62a} in their paramagnetic regime. However, until now, no results or arguments allowing a discrimination between these scenarios were reported. Thus, the origin for the observability of the ESR signal in the these two compounds remained a mystery.\\  
Recently, we have shown the CeTPO (T=Ru, Os) compound series to be an interesting class of Ce-based Kondo lattice systems at the border between intermetallic and oxide materials, presenting a layered structure and strong tendency to ferromagnetism.\cite{krellner07a} For CeRuPO a pronounced decrease of the electrical resistivity at temperatures below 50~K indicates the onset of coherent Kondo scattering which is confirmed by an enhanced thermopower. The reference compound LaRuPO neither shows signatures of the Kondo effect nor of magnetism of the Ru atoms.\cite{krellner07a} The temperature and magnetic field dependence of magnetic susceptibility and specific heat evidence ferromagnetic (FM) order at $T_{\rm C}=15\,$K. Thus, CeRuPO ($T_{\rm K}\simeq10\,$K) seems to be a rare example of a $f$-electron based FM Kondo lattice \cite{krellner07a}, along further $d$-electron based Kondo lattice systems \cite{b}. In contrast, CeOsPO shows antiferromagnetic (AFM) order at $T_{\rm N}=4.4\,$K, despite only minor changes in lattice parameters and electronic configuration. Therefore, CeRuPO and CeOsPO are ideally suited to study the difference between FM and AFM Kondo lattices. 
We present ESR results on these new Ce-based systems and discuss the ESR of further Ce- or Yb- based compounds with either ferro- or antiferromagnetic exchange and different strength of the Kondo interaction. The results demonstrate that the above mentioned unexpected ESR observation in heavy fermion metals is restricted neither to Yb-based compounds nor to compounds close to a quantum critical point, nor to compounds with a strong Kondo interaction. Instead, they indicate that the observability of the ESR signal in a dense Kondo lattice is connected with the presence of FM correlations between the Kondo ions.\\
%
%
We used polycrystalline samples of CeTPO (T=Ru, Os) whose preparation and characterization was described in Ref.\cite{krellner07a}. X-ray powder diffraction and energy dispersive X-ray measurements confirmed the formation of single phase CeTPO. The tetragonal crystal structure ($P4/nmm$) contains alternating layers of OCe$_4$ and TP$_4$ tetrahedra. It is worth to point out that despite the presence of oxygen the compounds show pronounced metallic properties with a residual resistivity $\rho_0=1.5\,\mu\Omega{\rm cm}$ for CeRuPO.
%
%
ESR probes the absorbed power $P$ of a transversal magnetic microwave field as a function of an external, static magnetic field $B$. To improve the signal-to-noise ratio, we used a lock-in technique by modulating the static field, which yields the derivative of the resonance signal $dP/dB$. The ESR experiments were performed at X-band frequencies ($\nu\approx9.4$~GHz) 
and at temperatures between 5~K$\leq T\leq$ 125~K. The measured ESR spectra could be fitted with a Lorentzian lineshape. Here we discuss the lineshape parameters linewidth ($\Delta B$) and resonance field ($B_{\rm res}$) which determines an effective ESR $g$-factor $(g=h\nu/\mu_{\rm B}B_{\rm res}$). In case of a local ESR spin probe in a metallic environment $\Delta B$ measures the spin-spin relaxation rate of the local spin, whereas $g$ is determined both by the magnitude of the local moment and the static magnetic field at the spin site.\\  
Fig. \ref{fig1} shows the main result of our ESR measurements on CeTPO (T=Ru, Os) at temperatures just above the %
%
respective magnetic ordering. The ESR investigations of both compounds were performed with the same instrument parameters and almost the same sample masses. For CeRuPO, which shows FM correlations, a well defined and intense ESR signal could easily be observed whereas the AFM correlated CeOsPO is absolutely ESR silent, even when using the largest instrument amplification. The different nature of magnetic correlations in CeRuPO and CeOsPO seems to be the only difference which is relevant for the ESR observation. In fact, many of their properties are similar: the preparation method is identical, the lattice parameters are very close, and the disorder effect on the electrical resistivity is comparable.\\  
The dashed line in Fig.\ref{fig1} fits the CeRuPO spectrum with a metallic Lorentzian line shape (including the effects of the counter rotating
component of the linearly polarized microwave field). The observed asymmetry of the line indicates the metallic character of the sample powder
and reflects a penetration depth being smaller than the grain size \cite{feher55a}.\\ 
Fig. \ref{fig2} shows the temperature dependence of the ESR spectra which is distinctly different for temperatures above and below $T_{\rm C}$. This difference indicates that the observed ESR reflects the bulk magnetic properties of CeRuPO. Above  $T_{\rm C}$, the resonance field is strongly temperature dependent and varies between 310~mT at 15~K and 210~mT at 24~K. With decreasing temperature, when entering the FM phase, the ESR signal splits into a low field (LF) and high field (HF) component. Both show a pronounced temperature dependence of their resonance fields $B_{\rm res}^{\rm LF,HF}(T)$ which shift in opposite directions. The decrease of $B_{\rm res}^{\rm LF}(T)$ with decreasing temperature reflects the developing internal magnetic fields because of the onset of  FM order. The HF component is much less intense and gets visible below $\approx$13~K (see dashed line fit in Fig. \ref{fig2}). $B_{\rm res}^{\rm HF}(T)$ shifts to higher fields with decreasing temperature, continuing the temperature behavior of the resonance field at $T>T_{\rm C}$. This is unusual, since one would not expect the resonance field to increase when approaching and entering the FM ordered state. Instead it should inversely follow the temperature dependence of the magnetization (taken at the ESR X-band field) \cite{taylor75b}. Magnetization and preliminary ESR results on a CeRuPO single crystal show that this unusual temperature dependence is due to a complex anisotropy, with the single ion anisotropy being opposite to the exchange anisotropy.\cite{krellner07b} A detailed ESR study is going on and shall be presented in a dedicated forthcoming paper.\cite{foerster07a}
%
%
%
In the paramagnetic regime, at $T=20$~K, the resonance field corresponds to an effective ESR $g$ factor of 2.6. This value is close to the expected value from the $\Gamma_6$ wavefunction of a Ce$^{3+}$ ion ($J = 5/2$) in tetragonal point symmetry.\\
To the best of our knowledge an ESR signal due to dense Ce$^{3+}$ magnetic moments in an environment with conduction electrons has only been reported for the Kramers salt CeP \cite{huang76a}. There the Ce$^{3+}$ resonance originates from transitions within the excited $\Gamma_8$ quartet. Note, that although CeP shows AFM correlations with an anomaly in magnetic susceptibility at $T_{\rm N}\simeq9$~K FM interactions are indicated by a positive Weiss temperature. This is also reflected in the Curie-Weiss behavior of the ESR $g$ factor with a positive Weiss temperature. Several ESR investigations of diluted Ce$^{3+}$ in metals have been performed in the cubic monopnictides with a $\Gamma_7$ ground state multiplet \cite{sugawara76a, wienand82a}. It turned out that Ce$^{3+}$ ESR is hard to observe because of a small $\Gamma_7-\Gamma_8$ splitting and the concomitant large linewidth contribution from the excited CEF level.\\   
The inset of Fig.\ref{fig2} shows the temperature dependence of the linewidth below $T_{\rm C}$ (solid symbols: LF component; open circles: HF component) and above $T_{\rm C}$ where only one single, well defined line (see line-fit in Fig.\ref{fig1}) could be observed. 
Within the temperature range where the HF component can be observed its linewidth changes only weakly whereas the linewidth of the LF component shows an extreme broadening towards low temperatures. We therefore suspect that the two components refer to different in-plane and out-of-plane magnetic properties of CeRuPO. This anisotropy becomes apparent below $T_{\rm C}$ and thus the ESR signal of polycrystalline samples splits. Within the FM phase no hysteresis effects could be observed in the ESR spectra. This absence is probably due to the weakness of the hysteresis effect in magnetization measurements at 2~K of oriented powder \cite{krellner07a}.\\
The strong linewidth increase of the LF component when passing $T_{\rm C}$ is typical for the onset of magnetic ordering of the ESR probe spins: the spin fluctuation time increases, the exchange narrowing mechanism gets less effective and inhomogeneous broadening gains influence, leading to a strong increase of the linewidth. The relatively sudden linewidth increase immediately below $T_{\rm C}$ indicates that the effect of atomic disorder or short range magnetic order is negligible for the ESR linewidth. A broadening which occurs already at temperatures far above $T_{\rm C}$ was attributed to effects of atomic disorder in various rare earth containing FM intermetallic compounds \cite{taylor75b}. Therefore, the observed $\Delta B(T)$ behavior indicates that the investigated CeRuPO sample is a ferromagnet with a high grade of homogeneity.\\ 
At temperatures above $T_{\rm C}$, $\Delta B(T)$ reflects the typical ESR behavior of a magnetic moment in a metallic host \cite{barnes81a}. $\Delta B(T)$ increases linearly with $\simeq 10$~mT/K in a small region up to $\approx18$~K whereas at higher temperatures an exponential increase dominates. A small amplitude and a large linewidth makes the signal undetectable for $T\gtrsim24$~K. A very similar behavior of $\Delta B(T)$ is found for diluted Ce$^{3+}$ in cubic LaAs, for instance \cite{wienand82a}.\\ 
%
%
%
%
Our comparison of the ESR properties of CeTPO (T=Ru,Os) provides a first evidence for a connection between the observability of the ESR line and the presence of ferromagnetic correlations. We obtained more experimental arguments for this conclusion by investigating further Yb or Ce based intermetallic compounds with either FM or AFM correlations and different strength of the Kondo interaction (see Tab. \ref{tab1}). Besides the positive ESR observations in YbRh$_{2}$Si$_{2}$ and 
\begin{table*}
\caption{\label{tab1}Intermetallic compounds with predominating AFM or FM spin correlations investigated by X-band ESR. ''Kondo" denotes a Kondo lattice behavior in $\rho(T)$ or other properties.}
\begin{ruledtabular}
\begin{tabular}{ccccc}
Compound & AFM & FM & Kondo & ESR signal\\ \hline
CeRuPO & - & \checkmark & \checkmark & Yes \\
CeOsPO & \checkmark & -  & \checkmark & No \\
YbRh & - & \checkmark & - & Yes \\
YbRh$_{2}$Si$_{2}$ \cite{trovarelli00a} & - & \checkmark & \checkmark &Yes \cite{sichelschmidt03a}\\
YbIr$_{2}$Si$_{2}$ (I-type) \cite{hossain05a} & - & \checkmark & \checkmark & Yes  \cite{sichelschmidt07a} \\
YbIr$_{2}$Si$_{2}$ (P-type) \cite{hossain05a} & \checkmark & - & \checkmark & No \\
Yb$_4$Rh$_{7}$Ge$_{6}$ \cite{ferstl04a} & \checkmark & - & - & No \\
YbNi$_2$B$_2$C \cite{bonville99a} & \checkmark & - & \checkmark & No \\
CeCu$_2$Si$_2$ (S/A) & \checkmark & - & \checkmark & No \\
CeNi$_2$Ge$_2$ & \checkmark & - & \checkmark & No \\
CeCu$_{\rm 6-x}$Au$_{\rm x}$ (x=0, 0.1) & \checkmark & - & \checkmark & No \\
\end{tabular}
\end{ruledtabular}
\end{table*}
YbIr$_{2}$Si$_{2}$ (I-type) \cite{sichelschmidt03a,sichelschmidt07a} which both exhibit strong FM correlations (as indicated, e.g., by an enhanced Sommerfeld-Wilson ratio) \cite{gegenwart05a,hossain05a} we found an ESR signal in yet another Yb-based compound with FM correlations, YbRh. However, in contrast to both above mentioned heavy fermion compounds, Kondo-type features are neither found in the electrical resistivity nor in the specific heat of YbRh. FM order of Yb$^{3+}$ magnetic moments below $T_{\rm C}=1.2$~K is confirmed by magnetic susceptibility and magnetization measurements as well as by the field dependence of the ordering temperature. \cite{Jeevan07} We investigated polycrystalline YbRh (CsCl structure) and observed a clean and well defined ESR signal within our accessible temperature range 3-30~K. The linewidth (minimum $\Delta B=220$~mT at $T=3$~K) shows a temperature dependence typical for local moments in metals. The ESR $g$-value reaches a constant value of 2.55 for $T\ge10$~K which is comparable to the $g$-values for Yb$^{3+}$ in cubic monopnictides \cite{wienand82a}.\\      
We looked for an ESR line in further Ce- and Yb-based compounds, with different strength of the Kondo interaction, but all with dominant antiferromagnetic correlations. Although the measurements were performed on high quality single crystals in a wide temperature range around $T_{\rm K}$ and / or the magnetic ordering temperature, we did not observe an ESR signal in any of these systems (see Tab. \ref{tab1}).  
P-type tetragonal YbIr$_{2}$Si$_{2}$ \cite{hossain05a} and cubic Yb$_4$Rh$_{7}$Ge$_{6}$ \cite{ferstl04a}, which both have been grown from In-flux as YbRh$_{2}$Si$_{2}$, display AFM ordering and exhibit more stable Yb$^{3+}$ moments than YbRh$_{2}$Si$_{2}$ (because of a weaker Kondo interaction). 
CeCu$_{2}$Si$_{2}$ (S/A type), CeNi$_{2}$Ge$_{2}$ and CeCu$_{\rm 6-x}$Au$_{\rm x}$ (x=0, 0.1) are three well established heavy fermion systems close to a quantum critical point. However, the former two are expected to present a spin density wave critical point, while the latter is suspected to present a local quantum critical point.\\  
Our results demonstrate that the observability of the ESR line in a dense Kondo lattice system is neither connected with specific properties of Yb or Ce, nor to the strength of the Kondo interaction, nor to the proximity of a (quantum) critical point (and thus also not to the character of the critical point). In contrast, these results give a strong evidence for the importance of ferromagnetic correlations for the narrowing of the ESR line. This conclusion bears a strong analogy to the situation in itinerant transition metal compounds. Itinerant $d$-systems with FM correlations may show a strong enhancement of their conduction electron susceptibility. This enhancement leads to an enhanced shift of the ESR $g$-value as well as a reduced Korringa broadening of the linewidth \cite{peter67a}. As a consequence an ESR signal may even be observed from the conduction electrons in the form of a collective conduction spin ESR (CESR) with a linewidth which is narrowed by electron-electron exchange interactions.\cite{fredkin72a}
Various examples of CESR in enhanced spin-susceptibility metals have been reported: Pd \cite{monod78a} and TiBe$_{2}$ \cite{shaltiel87a} which both show enhanced Sommerfeld-Wilson ratios, and the weak itinerant ferromagnet ZrZn$_{2}$ \cite{walsh70a}. However, we are not aware of any example of a strongly anisotropic CESR and an analogy to the ESR in Kondo lattice systems is not straightforward. Phenomenologically, the anisotropy may arise from the strong coupling between the local magnetic moments and the conduction electrons while a "bottleneck" relaxation framework may apply as indicated from ESR experiments on Yb$_{1-{\rm x}}$La$_{\rm x}$Rh$_{2}$Si$_{2}$.\cite{wykhoff07a}\\
%
%
%
We conclude that a substantial amount of experimental data indicates the importance of strong FM correlations among the ESR probed magnetic moments for the observability of an ESR signal in Kondo lattice systems in general. This conjecture is most strongly supported by the presence of an ESR signal in CeRuPO only since the dominant magnetic interaction is switched from AF (CeOsPO) to FM (CeRuPO) while keeping structural properties and disorder effects practically unchanged. These results not only provide a firm experimental basis for the understanding of the unexpected ESR line in YbRh$_{2}$Si$_{2}$ \cite{sichelschmidt03a} and YbIr$_{2}$Si$_{2}$ \cite{sichelschmidt07a}, but more importantly establish ESR as a new experimental tool for a direct study of the Kondo ion in dense Kondo systems. Since NMR and NQR cannot be performed on Ce and were yet not successful on Yb systems with strong magnetic correlations, our results establish ESR as a unique local probe for a direct investigation of the spin dynamics of the Kondo ion. The standard ESR technique (X-band) is especially qualified for this purpose because the magnetic field ($\ll 1$~T) is small enough to leave the Kondo state undisturbed. Preliminary investigations on the field and temperature dependence of the ESR response in YbRh$_{2}$Si$_{2}$ evidence e.g. a correlation between the evolution of the $g$-factor and the evolution of thermodynamic properties \cite{remark1}. Such investigations should give a deeper insight in the way the heavy quasiparticles are formed upon lowering the temperature or the field. \\  
%
%
%
We would like to thank J. Ferstl, M. Deppe, and O. Stockert for providing us single crystals of Yb$_4$Rh$_{7}$Ge$_{6}$, CeNi$_2$Ge$_2$, and CeCu$_{\rm 6-x}$Au$_{\rm x}$ (x=0, 0.1), respectively.
%

\begin{thebibliography}{21}
\bibitem{remark1}
High field ESR experiments of YbRh$_{2}$Si$_{2}$ are presently under way.
\end{thebibliography}

\end{document}